\documentclass[aps,pra,twocolumn,letterpaper,groupedaddress,10pt]{revtex4-1} 
\usepackage{amsmath,amssymb}
\usepackage{graphicx,mathptmx}
\usepackage[pdftex,dvipsnames,usenames]{xcolor}
\usepackage[colorlinks=true,urlcolor=blue,citecolor=blue,linkcolor=blue]{hyperref}

\begin{document}
\title{Wave-particle duality revisited: Neither wave nor particle}

\author{Jan Sperling}
	\affiliation{Integrated Quantum Optics Group, Applied Physics, University of Paderborn, 33098 Paderborn, Germany}

\author{Syamsundar De}
	\affiliation{Integrated Quantum Optics Group, Applied Physics, University of Paderborn, 33098 Paderborn, Germany}

\author{Thomas Nitsche}
	\affiliation{Integrated Quantum Optics Group, Applied Physics, University of Paderborn, 33098 Paderborn, Germany}

\author{Johannes Tiedau}
	\affiliation{Integrated Quantum Optics Group, Applied Physics, University of Paderborn, 33098 Paderborn, Germany}

\author{Sonja Barkhofen}
	\affiliation{Integrated Quantum Optics Group, Applied Physics, University of Paderborn, 33098 Paderborn, Germany}

\author{Benjamin Brecht}
	\affiliation{Integrated Quantum Optics Group, Applied Physics, University of Paderborn, 33098 Paderborn, Germany}

\author{Christine Silberhorn}
	\affiliation{Integrated Quantum Optics Group, Applied Physics, University of Paderborn, 33098 Paderborn, Germany}

\date{\today}

\begin{abstract}
	A textbook interpretation of quantum physics is that quantum objects can be described in a particle or a wave picture, depending on the operations and measurements performed.
	Beyond this widely held believe, we demonstrate in this contribution that neither the wave nor the particle description is sufficient to predict the outcomes of quantum-optical experiments.
	To show this, we derive correlation-based criteria that have to be satisfied when either particles or waves are fed into our interferometer.
	Using squeezed light, it is then confirmed that measured correlations are incompatible with either picture.
	Thus, within one single experiment, it is proven that neither a wave nor a particle model explains the observed phenomena.
	Moreover, we formulate a relation of wave and particle representations to two incompatible notions of quantum coherence, a recently discovered resource for quantum information processing.
	For such an information-theoretic interpretation of our method, we certify the nonclassicality of coherent states---the quantum counterpart to classical waves---in the particle picture, complementing the known fact that photon states are nonclassical in the typically applied wave picture.
\end{abstract}

\maketitle



	Particles and waves are mutually exclusive concepts in classical physics.
	A remarkable property of quantum theory is that such a distinction is no longer true.
	Consequently, the paradigm of wave-particle dualism was postulated, stating that quantum-mechanical objects can behave as waves or particles in different scenarios \cite{B28,AI38}.
	Today, we know that, for example, particles can indeed interfere \cite{DG28}, and waves can be observed as single quanta \cite{KDM77,GRA86}.
	Maybe to avoid giving up classical concepts, the popular interpretation still is that, depending on the context, quantum systems can be described in a particle or wave model; we aim at disproving this believe.

	While appearing as a counter\-intuitive concept, quantum mechanics and its experimental confirmations testify the validity of this theory.
	For example, electromagnetic light waves do come in quantized energy packets \cite{P00,E05}, photons.
	This seminal finding---together with the invention of the laser---mark the advent of quantum optics, the modern field of research devoted to the study of quantized radiation fields \cite{MW95,VW06,A12}.
	Beyond the investigation of fundamental aspects of nature, nowadays, optical quantum systems also provide a versatile platform for the implementation of novel applications in quantum computation and communication \cite{KLM01,KMNRDM07,GT07,S09}.

	Depending on the circumstances, the interpretation of experiments with light requires different descriptions.
	On the one hand, the vast field of classical optics is determined through Maxwell's theory, highlighting the wave nature of light.
	On the other hand, photon-antibunching \cite{KDM77} and Hong-Ou-Mandel (HOM) experiments \cite{HOM87} are key examples for observations which rely on the particle characteristics of light.
	Therefore, each experiment in quantum optics requires a different reference frame, favoring either a photon-particle description or electromagnetic waves.
	However, it has not been challenged weather there are experimental situations in which either framework fails to explain the measured data.
	Among other reasons, this surprising gap in our knowledge may be due to a lack of criteria which are capable to simultaneously quantify the inconsistency with wave and particle models.
	Here, we overcome this deficiency.


\begin{figure}[b]
	\includegraphics[width=\columnwidth]{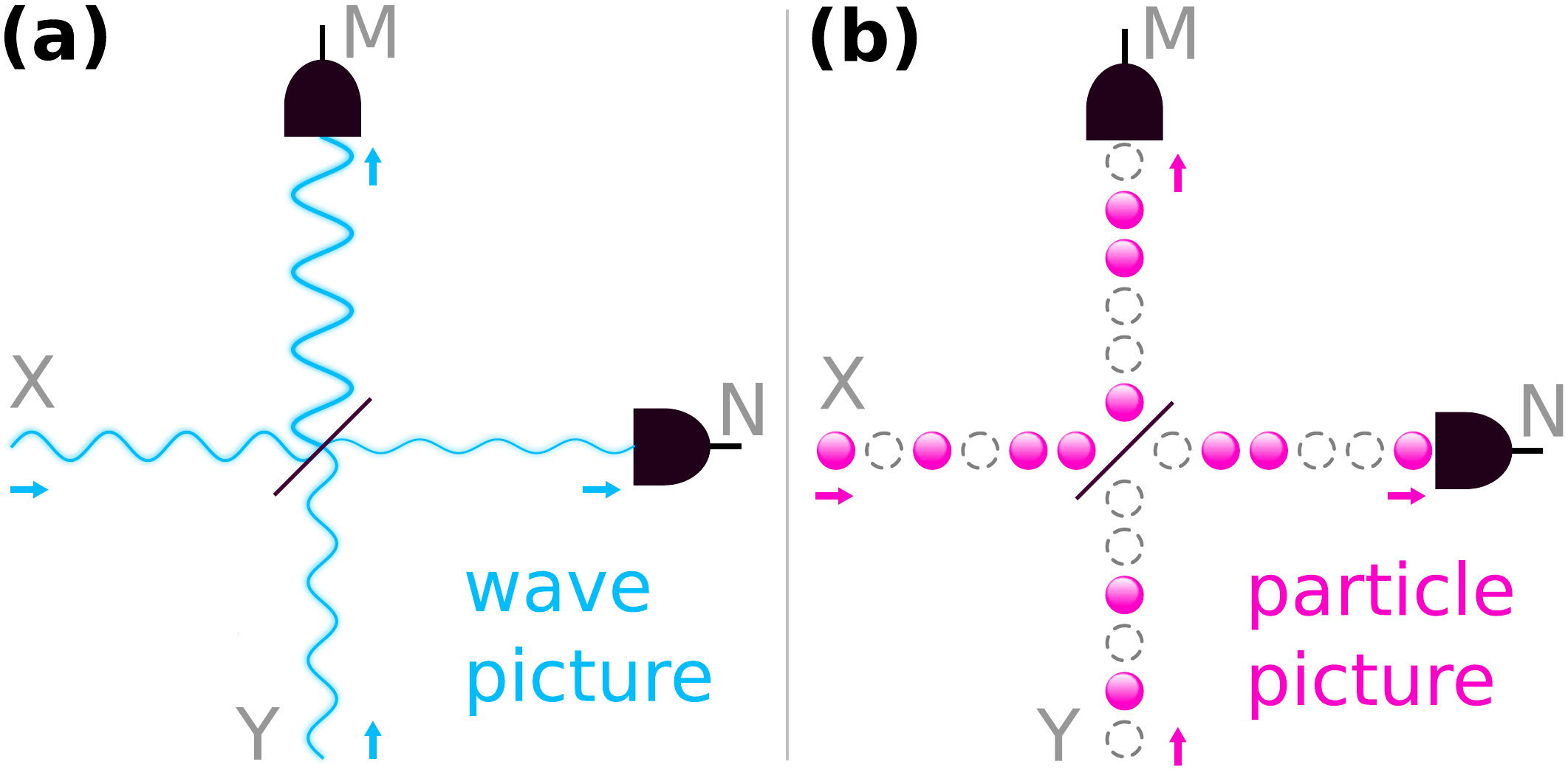}
	\caption{
		(a) Two waves interfere constructively (top) and destructively (right) on a 50:50 beam splitter, and these signals are then measured.
		(b) Incident particles to our device are distributed uniformly to one of the outputs and counted.
	}\label{fig:Outline}
\end{figure}

	For achieving our goal of demonstrating that neither waves nor particles deliver a viable model for quantum-optical experiments, we design a setup (see Fig. \ref{fig:Outline}) that is surprisingly simple and consists of building blocks which are known to be consistent with one of the classical pictures only.
	Specifically, we employ (50:50) beam splitters, whose operation is described in terms of electromagnetic waves, and photon detectors, a particle counter.

	Our figure of merit to assess the applicability of the distinct models under study is given in terms of the measured correlation which can be expected for waves and particles.
	Suppose $M$ and $N$ are the random variables that describe the top and right measurement outcomes, respectively, of our interferometer in Fig. \ref{fig:Outline}.
	In addition, $P(X,Y)$ is the probability distribution for input parameters $X$ (left) and $Y$ (bottom).
	The measured covariance matrix takes the form
	\begin{equation}
		\mathbf{C}=\begin{pmatrix}
			\mathrm{Var}(M) & \mathrm{Cov}(M,N)
			\\
			\mathrm{Cov}(M,N) & \mathrm{Var}(N)
		\end{pmatrix},
	\end{equation}
	where ``Var'' and ``Cov'' denote the variance and covariance, respectively.
	This covariance matrix can be decomposed as $\mathbf{C}=\mathbf{B}+\mathbf{C}'$ by applying the statistical laws of total variances and covariances \cite{B69,S95}, which are based on conditional expressions for input settings $Z=(X,Y)$.
	Specifically,
	\begin{equation}
		\mathbf{B}=\left(\begin{smallmatrix}
			\mathrm{E}[\mathrm{Var}(M\mid Z)] & \mathrm{E}[\mathrm{Cov}(M,N \mid Z)]
			\\
			\mathrm{E}[\mathrm{Cov}(M,N \mid Z)] & \mathrm{E}[\mathrm{Var}(N \mid Z)]
		\end{smallmatrix}\right)
	\end{equation}
	is the mean value of a conditional covariance matrix, where $\mathrm{E}$ denotes the expectation value with respect to the input distribution $P$,
	and
	$\mathbf{C}'=\left(\begin{smallmatrix}
		\mathrm{Var}[\mathrm{E}(M\mid Z)] & \mathrm{Cov}[\mathrm{E}(M\mid Z),\mathrm{E}(N\mid Z)]
		\\
		\mathrm{Cov}[\mathrm{E}(M\mid Z),\mathrm{E}(N\mid Z)] & \mathrm{Var}[\mathrm{E}(N\mid Z)]
	\end{smallmatrix}\right)$
	is a covariance matrix of conditional mean values \cite{B69,S95}.
	The latter covariance matrix is always positive semidefinite, $\mathbf{C}'\geq0$.
	Consequently, one readily finds that
	\begin{equation}
		\mathbf{C}-\mathbf{B}\geq0
	\end{equation}
	holds true for any classical distribution $P$.
	In particular, the minimal eigenvalue $e$ of the $2\times 2$ matrix $\mathbf{C}-\mathbf{B}$ has to be nonnegative.
	In the following, we apply the classical particle and wave picture to determine $\mathbf{B}$ for both cases, resulting in $\mathbf{B}_\mathrm{part.}$ and $\mathbf{B}_\mathrm{wave}$.
	Then, a violation of the above constraints, i.e., $e_\mathrm{wave}<0$ and $e_\mathrm{part.}<0$, verifies the incompatibility of the measured data with the particular model under study.

	Firstly, we may formulate bounds to the covariance matrix for particles.
	Regardless of the input port, a single photon has the probability $1/2$ to be directed to the top or right detector in Fig. \ref{fig:Outline}(b) as we have a 50:50 splitting ratio.
	Consequently, if $X$ independent particles of light from the left and $Y$ from the bottom direction are distributed to the detectors, we directly get a binomial output distribution, $\binom{X+Y}{M}[1/2]^{M}[1/2]^{X+Y-M}$, for the detection of $M$ photons on the top and $N=X+Y-M$ photons on the right detector.
	Thus, the covariance matrix takes the well-known form
	$\left(\begin{smallmatrix}
		(M+N)/4 & -(M+N)/4
		\\
		-(M+N)/4 & (M+N)/4
	\end{smallmatrix}\right)$
	under the constraints of the single input setting $Z=(X,Y)$.
	When averaging over arbitrary inputs, we then get the correspondingly defined matrix-type bound
	\begin{equation}
		\label{eq:BoundPart}
		\mathbf{B}_\mathrm{part.}=\frac{\mathrm{E}(M+N)}{4}
		\begin{pmatrix}
			1 & -1 \\ -1 & 1
		\end{pmatrix},
	\end{equation}
	where $\mathrm{E}(M+N)$ is the mean total particle number.
	It is worth mentioning that losses can be treated as a statistical mixture of fewer particles sent into our device, which is already captured by the averaged expression for $\textbf{B}_\mathrm{part.}$ above.

	Secondly, a classical wave model can be derived as follows.
	Suppose $Z=(X,Y)$ correspond to the two amplitudes of the electromagnetic fields at the input, cf. Fig. \ref{fig:Outline}(a).
	Then, the classical operation of the beam splitter maps $X\mapsto X'=(X+e^{i\vartheta}Y)/\sqrt 2$ and $Y\mapsto Y'=(Y-e^{-i\vartheta}X)/\sqrt 2$, with a phase factor $e^{i\vartheta}$.
	A classical model for each detector can be achieved by assuming that the detector---say the one on the right---consists of many elements (e.g., $D$ atoms) which individually have the probability $p(X';D)$ to result in small signal (i.e., click) for the field amplitude $X'$ of the impinging light on that elements and for arbitrary light-matter interactions \cite{MW95}.
	This means that for signals from $N$ elements, one obtains the distribution $\binom{D}{N}p(X';D)^N[1-p(X';D)]^{D-N}$.
	In the limit of a macroscopic number of atoms, it is well-known that one approaches the Poisson distribution $e^{-\lambda(X')}\lambda(X')^N/N!$, with a mean value $\lim_{D\to\infty} Dp(X';D)=\lambda(X')=E(N|Z)$.
	Note that this detector is not affected by the other (orthogonal) output mode of the beam splitter, $Y'$;
	thus, we have independent statistics for both detectors, which are assumed to operate in the same manner.
	Consequently, the covariance for the single setting of input amplitudes $X$ and $Y$ reads as
	$\left(\begin{smallmatrix}
		\lambda(X') & 0
		\\
		0 & \lambda(Y')
	\end{smallmatrix}\right)$.
	Note that losses can be modelled through the detection probability $p$, likewise $\lambda$.
	Again, we may average over arbitrary inputs $Z$ to get the desired matrix-type bound for waves,
	\begin{equation}
		\label{eq:BoundWave}
		\mathbf{B}_\mathrm{wave}=
		\begin{pmatrix}
			\mathrm{E}(M) & 0 \\ 0 & \mathrm{E}(N)
		\end{pmatrix}.
	\end{equation}

	Based on classical models for our device, consisting of an inherently wave part (beam splitter) and particle-based components (detectors), we have now formulated two constraints, $\mathbf{C}-\mathbf{B}_\mathrm{part.}\geq0$ and $\mathbf{C}-\mathbf{B}_\mathrm{wave}\geq0$, which have to be satisfied by particles and waves, respectively.
	It is also worth emphasizing that both criteria are formulated in terms of measurable quantities only.
	Furthermore, efficiencies do not affect the overall form of our expressions, rendering it robust against unavoidable losses.
	Similarly to Bell's approach to derive his seminal inequality \cite{B64}, our criteria are based on the existence of a classical statistical description, $P(X,Y)$, of our experiment to account for the observations made and which can be violated by quantum physics.

\begin{figure}[t]
	\includegraphics[width=\columnwidth]{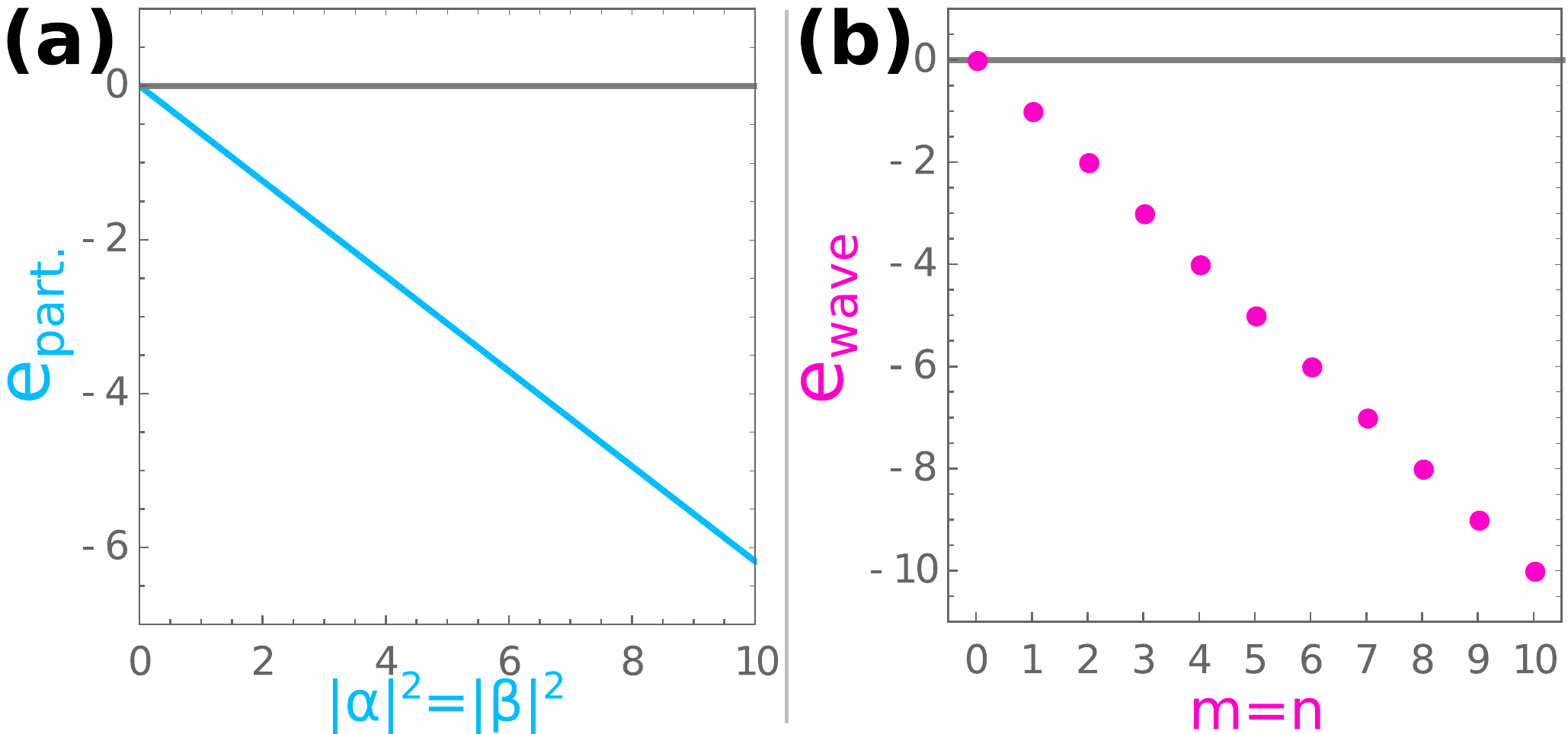}
	\caption{
		Theory benchmark for the lossless case.
		(a) Coherent states $|\alpha,\beta\rangle$ violate the particle criterion, $e_\mathrm{part.}< 0$, where $\alpha=\beta$ and $\vartheta=0$.
		(b) The photon-state input $|m,n\rangle$ for $m=n$ violates the wave nature of light, $e_\mathrm{wave}<0$, representing a generalized HOM experiment.
		The vacuum state $|0,0\rangle$ is classical in both pictures, $e_\mathrm{wave/part.}=0$.
	}\label{fig:Theory}
\end{figure}

	Thus, we may use quantum theory to make the predictions about the outcomes of our device for different input states; see the Supplemental Material (SM) \cite{Supplement} for details.
	Calculating the minimal eigenvalues $e_\mathrm{part.}$ of $\mathbf{C}-\mathbf{B}_\mathrm{part.}$ for photon-number input states $|m,n\rangle$, quantum particles of light, we confirm that this eigenvalue is nonnegative.
	Similarly, we find for the minimal eigenvalue of $\mathbf{C}-\mathbf{B}_\mathrm{wave}$ that $e_\mathrm{wave}\geq0$ holds true for coherent input states $|\alpha,\beta\rangle$, the quantum counterpart to classical waves.
	However, in Fig. \ref{fig:Theory}(a) and \ref{fig:Theory}(b), we can observe that $e_\mathrm{part.}<0$ for coherent states and $e_\mathrm{wave}<0$ for photons hold true, certifying the states' nonclassical properties in the corresponding antipodal pictures.
	Furthermore, we expect from our quantum-mechanical calculations that squeezed input states are promising candidates that violate both constraints simultaneously, motivating our implementation.


	We now briefly describe the experimental setup to realize the scheme in Fig. \ref{fig:Outline};
	see the SM \cite{Supplement} for all technical details.
	A titanium sapphire laser delivers picosecond pulses that pump a periodically poled potassium titanyl phosphate waveguide.
	This generates two-mode squeezed vacuum via type-II spontaneous parametric down-conversion.
	The two modes propagate through our fiber-loop-based time-multiplexing network \citep{NBKSSGPKJS18}, which yields two temporally overlapped squeezed vacuum states of orthogonal polarizations.
	Finally, to implement the protocol of Fig. \ref{fig:Outline}, the two single-mode squeezed states are symmetrically mixed using a half-wave plate followed by a polarizing beam splitter and detected using a two-mode time-multiplexed detection layout.

	The latter detection layout uses the same device for the two output pulses at different times, realized via a delay line \cite{SBVHBAS15} and satisfying the assumption that both detectors operate in the same manner.
	It is also worth emphasizing that true photon-number resolving detectors do not exist.
	Our detection scheme has indeed a pseudo-photon-number resolution which is rigorously accounted for by realizing small intensities only and adding a systematic error to our statistical analysis \cite{Supplement}.


\begin{figure}[t]
	\includegraphics[width=\columnwidth]{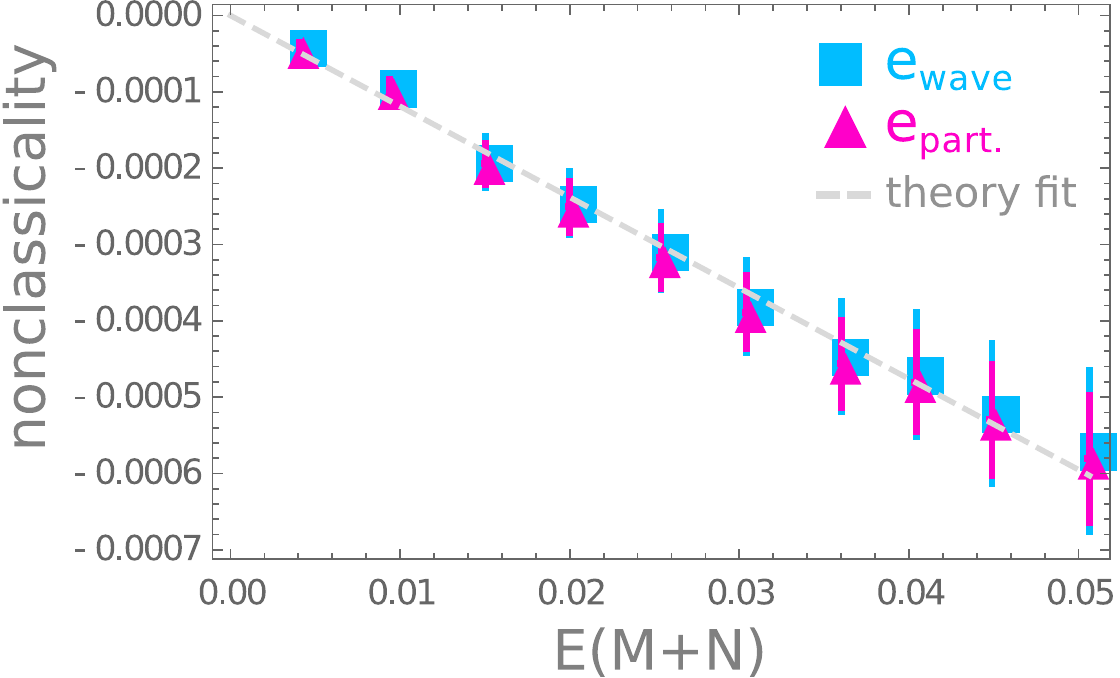}
	\caption{
		Experimentally verified wave-particle violation (nonclassicality $e_\mathrm{wave/part.}<0$) as a function of the total mean photon number $\mathrm{E}(M+N)$  (likewise, unit-free total intensity) for squeezed states.
		Classical wave and particle models predict values $e_\mathrm{part.}\geq0$ and $e_\mathrm{wave}\geq0$ and can be ruled out.
		However, quantum theory, not being restricted to waves or particles, is consistent with our data (dashed line).
		Note that the error bars for $\mathrm{E}(M+N)$, horizontal axis, are depicted but much smaller than the used triangle and square symbols.
	}\label{fig:Result}
\end{figure}

	In Fig. \ref{fig:Result}, we show the results of our correlation analysis when squeezed states enter our interferometric device.
	In particular, we can certify with high statistical significance---up to 8 standard deviations below the classical threshold of zero---that a nonclassical interpretation of our data is essential since neither the wave ($e_\mathrm{wave}<0$) nor the particle picture ($e_\mathrm{part.}<0$) is applicable.
	By contrast, quantum theory provides a correct description of our data which can be seen from the theory curve, dashed line in Fig. \ref{fig:Result}.

	The quantum-optical interpretation of our data goes as follows.
	When two squeezed states are combined on a beam splitter with specific phase and amplitude relations, an entangled two-mode squeezed vacuum state is produced which is ideally described through perfect photon-photon correlations.
	By contrast, recall that the particle model predicts anti-correlations and the wave model no correlations.
	Furthermore, our quantum-mechanical calculations \cite{Supplement} yield the linear relation $e_\mathrm{part.}=e_\mathrm{wave}=-\eta\mathrm{E}(M+N)/2$, which is in agreement to Fig. \ref{fig:Result} when taking error bars into account.
	The efficiency is the only fit parameter and thereby estimated as $\eta\approx2.4\%$, combining losses in the state generation, coupling, and detection part of our setup.
	This presents a challenging situation which is, however, demonstrated to not affect our conclusions of nonclassicality because of the robustness of our criteria.
	Still, a relatively long total measurement time ($\sim9\,\mathrm{min}$) for each squeezed state is required to accumulate enough data for a statistically significant evaluation.
	Furthermore, the conclusion that both the particle and wave description can be ruled out is verified for total intensities which span over one order of magnitude, $0.005\lessapprox\mathrm{E}(M+N)\lessapprox0.05$.


	Beyond considerations based on classical models, recent approaches, aimed at applications in quantum technology, sparked a renewed interest in questions about the wave-particle duality \cite{BBCH16,RPMAXSFS17,YHZZXLG18,DCPDS18}, even leading to revisiting more fundamental aspects of this duality \cite{QMLYN19,CBB19,QA19,QKMSVW19}.
	For instance, quantum key distribution protocols with coherent states (see, e.g., Refs. \cite{GG02,GAWBCG03}) requires a quantum-mechanical framework in which one has to evaluate the nonclassical character of such wave-like states as a quantum resource for this task.

	On the one hand, the traditional quantum-mechanical definition of a classical state, being consistent with the wave description, is based on the Glauber-Sudarshan representation \cite{G63,S63}.
	Here, a multimode state $\hat\rho$ is classical if
	\begin{equation}
		\label{eq:GSrep}
		\hat\rho={\int} d^2\alpha_1d^2\alpha_2\cdots P(\alpha_1,\alpha_2,\ldots)|\alpha_1,\alpha_2,\ldots\rangle\langle\alpha_1,\alpha_2,\ldots|
	\end{equation}
	and $P\geq 0$ for all coherent field amplitudes $(\alpha_1,\alpha_2,\ldots)$ hold true \cite{TG65,L86}.
	Clearly, a coherent state is classical in this regard.
	On the other hand, the recently developed resource theory for quantum coherence typically employs orthogonal states; see Refs. \cite{SAP17,CG18} for reviews.
	From this point of view, a state $\hat\rho$ is classical, i.e., incoherent, if it can be represented in a diagonal form of multimode photon-number states
	\begin{equation}
		\label{eq:QCrep}
		\hat\rho=\sum_{n_1,n_2,\ldots} P(n_1,n_2,\ldots) |n_1,n_2,\ldots\rangle\langle n_1,n_2,\ldots|.
	\end{equation}

	In Eqs. \eqref{eq:GSrep} and \eqref{eq:QCrep}, a classical state is given by a statistical mixture of tensor-product states which are either consistent with the wave picture, $|\alpha_1,\alpha_2,\ldots\rangle$, or particle description, $|n_1,n_2,\ldots\rangle$.
	It has been shown that if a state is nonclassical with respect to at least one of these complementary definitions, this state serves as a resource for quantum protocols, e.g., by allowing the generation of entangled states \cite{VS14,KSP16}.
	For additional discussions about the incompatibility of both concepts of nonclassicality, we refer to Refs. \cite{FP12,ASV13}.
	In connection to our criterion, we can further conclude that $e_\mathrm{wave}<0$ and $e_\mathrm{part.}<0$ serve as nonclassicality witnesses in quantum optics [Eq. \eqref{eq:GSrep}] and quantum information theory [Eq. \eqref{eq:QCrep}], respectively.
	This is true since---as shown with our theoretical analysis---the corresponding classical reference states, including mixtures thereof, cannot lead to negative eigenvalues.

	While the nonclassicality of photon-number states is well documented in the literature (see, e.g., the HOM experiment \cite{HOM87} and recent realizations for higher photon numbers \cite{Setal17}), an experimental verification of the nonclassicality of coherent states has, to our best knowledge, not been performed to date.
	For this reason, we prepared coherent states by coupling laser light into our fibre-loop time-multiplexing network instead of down-converted light.
	This prepares two orthogonally polarized coherent states to be probed with our device in Fig. \ref{fig:Outline}.

	The results of our analysis can be found in Table \ref{tab:Coherent}.
	Firstly, as expected, the data are consistent with the wave picture since deviations from $e_\mathrm{wave}=0$ are insignificant within the relative error margin.
	Secondly, however, we also clearly see that coherent states are nonclassical with respect to a reference frame based on particles, $e_\mathrm{part.}<0$.
	With increasing coherent amplitudes, determined through $\mathrm{E}(M+N)$, the certified nonclassicality is shown to increase, cf. Fig. \ref{fig:Theory}(a).
	Thus, the nonclassicality of coherent states is experimentally certified, implying the usefulness of such states for quantum information tasks.
	This also clarifies in a rigorous manner the frequently observed confusion between different communities that attach different definitions to what can be considered as nonclassical.

\begin{table}[t]
	\caption{
		Results, including relative errors, of the wave-particle classification for the generated coherent states with different mean total photon numbers (first column).
		The second column shows that $e_\mathrm{wave}=0$ holds true within error margins, which is the expected outcome for coherent states that are the quantum analog to classical waves.
		For most coherent amplitudes, the third column certifies with high statistical significance that coherent states are nonclassical with respect to the particle picture, $e_\mathrm{part.}<0$.
	}\label{tab:Coherent}
	\begin{tabular}{ccc}
		\hline\hline
		$\mathrm{E}(M+N)$ & $e_\mathrm{wave}$ & $e_\mathrm{part.}$ \\
		\hline
		$0.395\!{\times}\! 10^{-2}(1{\pm}0.006)$ & $-1\!{\times}\! 10^{-6}(1{\pm}15)$ & $-0.004\!{\times}\! 10^{-3}(1{\pm}3)\phantom{.000}$ \\
		$0.777\!{\times}\! 10^{-2}(1{\pm}0.004)$ & $-1\!{\times}\! 10^{-6}(1{\pm}19)$ & $-0.247\!{\times}\! 10^{-3}(1{\pm}0.063)$ \\
		$1.175\!{\times}\! 10^{-2}(1{\pm}0.003)$ & $-6\!{\times}\! 10^{-6}(1{\pm}\phantom{0}7)$ & $-0.660\!{\times}\! 10^{-3}(1{\pm}0.029)$ \\
		$2.078\!{\times}\! 10^{-2}(1{\pm}0.003)$ & $-4\!{\times}\! 10^{-7}(1{\pm}88)$ & $-1.824\!{\times}\! 10^{-3}(1{\pm}0.014)$ \\
		$3.025\!{\times}\! 10^{-2}(1{\pm}0.003)$ & $\phantom{-}5\!{\times}\! 10^{-7}(1{\pm}77)$ & $-3.178\!{\times}\! 10^{-3}(1{\pm}0.010)$ \\
		$3.937\!{\times}\! 10^{-2}(1{\pm}0.003)$ & $\phantom{-}1\!{\times}\! 10^{-6}(1{\pm}38)$ & $-4.432\!{\times}\! 10^{-3}(1{\pm}0.009)$ \\
		$4.482\!{\times}\! 10^{-2}(1{\pm}0.003)$ & $-4\!{\times}\! 10^{-5}(1{\pm}\phantom{0}4)$ & $-5.302\!{\times}\! 10^{-3}(1{\pm}0.008)$ \\
		\hline\hline
	\end{tabular}
\end{table}


	In summary, we revisited the wave-particle duality.
	Specifically, we have shown in theory and experiment that, already for relatively simple instances of quantum-optical setups, a particle and wave interpretation of quantum light simultaneously fails to explain the measured data.
	This proves that neither the wave nor the particle description is sufficient to explain the properties of light which, consequently, leads to questioning the usefulness of such classical notions in the context of quantum systems.

	For showing this, we derived criteria which have to be satisfied if either picture provides a valid description of our interference device.
	This setup was purposefully designed in such a way that it combines components which are naturally connected to only one interpretation of quantum light, i.e., using beam splitters for waves and photon counters for particles.
	Feeding our device with squeezed states, our analysis showed with high statistical significance that the observed correlations are inconsistent with both the particle and wave model at the same time.
	However, the quantum interpretation provides a correct model of our observation as it is not limited to either model.
	We emphasize that all results are directly obtained from the data taken, without performing corrections or postselections and including careful estimates of random and systematic uncertainties; that our criteria are robust against losses and derived from classical considerations; and that the applied method extends to higher orders and multimode systems via multivariate and higher-order conditional cummulants.

	Furthermore, we realized a scenario which employs coherent states.
	In this case, our criteria showed that the particle description could be ruled out conclusively, demonstrating the nonclassicality of coherent states.
	This complements the predominantly applied viewpoint in which it has been frequently shown that photon-number states are nonclassical with respect to a wave picture.
	The verified nonclassicality also enables us to assess the usefulness of coherent states as a resource in information processing, inspiring an alternative way of thinking about the wave-particle duality as a quantum resource.

	In the future, one has to expect a high demand for protocols which simultaneously require multiple resources of quantum coherence for developing interfaces between different quantum information platforms.
	Our approach, based on complementary waves and particle states, may serve as a starting point for such developments.
	In this context, our results on squeezed states confirm their simultaneous nonclassicality with respect to two complementary notions of quantum coherence.
	This implies the potential to use such states for interconnecting different resources in quantum technologies.

	In conclusion, we developed an alternative view on a fundamental paradigm of quantum physics by rejecting both the wave and particle concept in a single experiment and by exploring the resulting usefulness as a resource for quantum information applications.


	 The Integrated Quantum Optics group acknowledges financial support from the Gottfried Wilhelm Leibniz-Preis (Grant no. SI1115/3-1) and the European Commission with the ERC project QuPoPCoRN (Grant no. 725366).


\appendix
\onecolumngrid
\clearpage
\section*{Supplemental Material}

\begin{quote}
	Here, we provide technical details about
	the modeling of the quantum states under study, including losses (Sec. \ref{sec:quantumstates}),
	the data analysis and error estimation (Sec. \ref{sec:dataanalysis}),
	as well as our experimental setup (Sec. \ref{sec:experiement}).
\end{quote}

\twocolumngrid

\section{Quantum description}\label{sec:quantumstates}

	To interpret our approach in the quantum-mechanical domain, we can straightforwardly identify measured moments.
	In particular, we have $\mathrm{E}(M)=\langle\hat n_A\rangle$ and $\mathrm{E}(N)=\langle \hat n_B\rangle$ for the first moments, with $\hat n_S$ denoting the photon-number operator of the measured mode $S\in\{A,B\}$.
	The variances and covariance are defined similarly, $\mathrm{Var}(M)=\langle(\Delta\hat n_A)^2\rangle$, $\mathrm{Var}(N)=\langle(\Delta\hat n_B)^2\rangle$, and $\mathrm{Cov}(M,N)=\langle(\Delta\hat n_A)(\Delta\hat n_B)\rangle$, where $\Delta\hat x=\hat x-\langle\hat x\rangle$ for arbitrary observables $\hat x$.
	In addition, loss can be described via the standard transformations \cite{VW06,A12} $\langle \hat n_S\rangle\mapsto\eta\langle \hat n_S\rangle$, $\langle \hat n_S^2\rangle\mapsto\eta^2\langle \hat n_S^2\rangle+\eta(1-\eta)\langle\hat n_S\rangle$, and $\langle \hat n_A\hat n_B\rangle\mapsto\eta^2\langle \hat n_A\hat n_B\rangle$, for a quantum efficiency $0\leq\eta\leq1$.

	Interfering two identical squeezed states on a 50:50 beam splitter with a well-adjusted phase relation yields a two-mode squeezed vacuum state, which has the joint photon-number statistics, $p_{n_A,n_B}=\delta_{n_A,n_B} (1-q)q^{n_A}$, where $0<q<1$ corresponds to the amount of squeezing and $\delta$ denotes the Kronecker symbol.
	From this distribution, we can directly compute ideal moments as
	\begin{equation}
		\langle \hat n_A\rangle {=} \langle \hat n_B\rangle
		{=} \frac{q}{1{-}q}
		\text{ and }
		\langle \hat n_A^2\rangle {=} \langle \hat n_B^2\rangle {=} \langle \hat n_A\hat n_B\rangle
		{=} \frac{q(1{+}q)}{(1{-}q)^2}.
	\end{equation}
	Including a non-unit quantum efficiency, $0<\eta<1$ for $A$ and $B$, we then get the covariance matrix,
	\begin{equation}
		\mathbf{C} = \frac{\eta^2q}{(1-q)^2}
		\begin{pmatrix}1&1\\1&1\end{pmatrix}
		+\frac{\eta(1-\eta)q}{1-q}
		\begin{pmatrix}1&0\\0&1\end{pmatrix},
	\end{equation}
	as well as the matrix-valued bounds
	\begin{equation}
		\mathbf{B}_\mathrm{wave}=
		\frac{\eta q}{1-q}
		\begin{pmatrix}1&0\\0&1\end{pmatrix}
		\text{ and }
		\mathbf{B}_\mathrm{part.}=
		\frac{\eta q}{2(1-q)}
		\begin{pmatrix}1&-1\\-1&1\end{pmatrix}.
	\end{equation}
	From this, we compute the minimal eigenvalue for $\mathbf{C}-\mathbf{B}_\mathrm{wave}$, $e_\mathrm{wave}=-\eta^2 q/(1-q)$.
	Similarly, we find the same eigenvalue $e_\mathrm{part.}=-\eta^2 q/(1-q)$ for $\mathbf{C}-\mathbf{B}_\mathrm{part.}$.
	The total mean photon number is $\langle \hat n_A+\hat n_B\rangle=2\eta q/(1-q)$, including losses.
	Note that the eigenvalue is negative for all non-zero efficiencies, and we can identify $e_\mathrm{wave}=e_\mathrm{part.}=(-\eta/2)\langle \hat n_A+\hat n_B\rangle$.
	This relation is used for the fit of the data to this quantum-mechanical prediction, cf. Fig. 3 in the main part, and in which we have only one free parameter, $\eta$. 

	Analogously, one can compute the corresponding eigenvalues for other input states.
	In particular, when including losses, we find for the coherent state $|\alpha,\beta\rangle$
	\begin{equation}
	\label{eq:CoherentStateTheory}
	\begin{split}
		e_\mathrm{wave} = & 0 \quad\text{and}
		\\
		e_\mathrm{part.} = & \frac{\eta(|\alpha|^2+|\beta|^2)}{4}\left[
			1 {-} \sqrt{1 {+} \left(\frac{4\mathrm{Re}[e^{i\vartheta}\beta\alpha^\ast]}{|\alpha|^2+|\beta|^2}\right)^2}
		\right],
	\end{split}
	\end{equation}
	and for the photon-number state $|m,n\rangle$
	\begin{equation}
	\label{eq:FockStateTheory}
	\begin{split}
		e_\mathrm{wave} = & -\eta^2\frac{m+n}{2} \quad\text{and}
		\\
		e_\mathrm{part.} = & \min\left\{
			\eta^2 mn,\eta(1-\eta)\frac{m+n}{2}
		\right\}.
	\end{split}
	\end{equation}
	Recall that the beam splitter transformation maps annihilation operators as $\hat a\mapsto (\hat a+e^{i\vartheta}\hat b)/\sqrt 2$ and $\hat b\mapsto (-e^{-i\vartheta}\hat a+\hat b)/\sqrt 2$.

\begin{figure}[b]
	\includegraphics[width=\columnwidth]{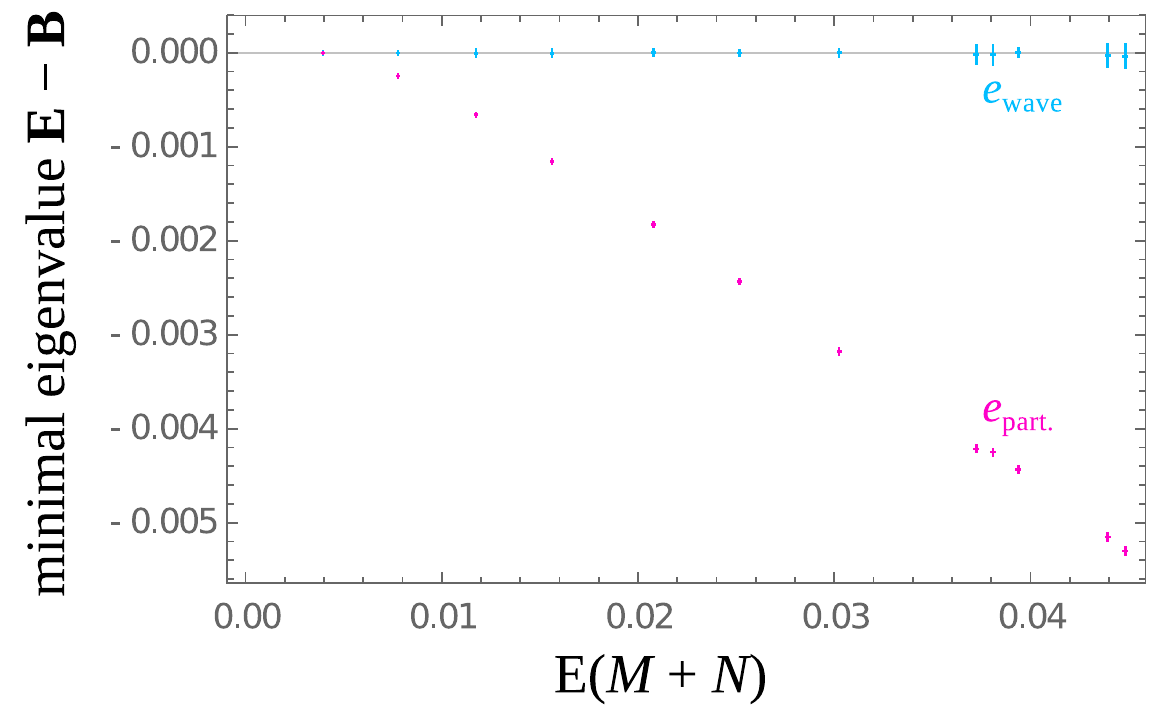}
	\caption{
		For coherent states, our results (including uncertainties, one standard deviation plus systematic error [cf. Sec. \ref{sec:dataanalysis}]) are shown, discussed in Table I of the main text and including additional data points not shown there. 
		The results are consistent with the expected value zero for waves, $e_\mathrm{wave}\approx 0$ (cyan) [cf. Eq. \eqref{eq:CoherentStateTheory}].
		The particle description is clearly not applicable, $e_\mathrm{part.}<0$ (magenta).
	}\label{fig:coherent}
\end{figure}

	The above results for the special case $\eta=1$ (as well as $m=n$ and $\alpha=\beta$, with $\vartheta=0$) are used for the plots in Fig. 2 in the main text. 
	Here, the results of our data analysis for coherent states (see Table I in the main text) are additionally visualized in Fig. \ref{fig:coherent}. 
	Furthermore, one can rather straightforwardly deduce from Eqs. \eqref{eq:CoherentStateTheory} and \eqref{eq:FockStateTheory} that $e_\mathrm{part.}$ for coherent states and $e_\mathrm{wave}$ for photon-number states are negative for $\eta>0$ iff the same holds true for the lossless case.

	In general, losses in optical systems can be modeled via a beam splitter operation in which the quantum efficiency $\eta$ corresponds to the amount of transmitted light, while also tracing over the loss (reflected) mode \cite{VW06,A12}.
	On the one hand, one gets $|\alpha\rangle\mapsto |\sqrt\eta\alpha\rangle$ for coherent light.
	On the other hand, the loss relation for an $n$-photon state reads $|n\rangle\langle n|\mapsto\sum_{m=0}^n\binom{n}{m}\eta^m(1-\eta)^{n-m}|m\rangle\langle m|$.
	In both cases, we obtain (mixed) states which are still classical within their respective reference frames, explaining the robustness of our method against losses.

	Let us give some additional examples of states which are interesting for our considerations.
	It is worth recalling that we study two complementary notions of quantum coherence, one defined as a convex mixture of coherent states---the typical approach in quantum optics---and one as statistical ensembles of photon-number states---more common in quantum information theory.
	A state which is classical in the corresponding wave and particle picture is a thermal state,
	\begin{equation}
		\label{eq:ThermalState}
		\int d^2\alpha\,\frac{e^{-|\alpha|^2/\bar n}}{\pi\bar n}|\alpha\rangle\langle\alpha|
		=\frac{1}{1+\bar n}\sum_{n=0}^\infty \left[\frac{\bar n}{\bar n +1}\right]^n|n\rangle\langle n|.
	\end{equation}
	The same holds true for the vacuum state $|0\rangle\langle 0|$.
	When probed with our interferometer, thermal (vacuum) states at each input result in $e_\mathrm{wave}\geq0$ and $e_\mathrm{part.}\geq0$, which is clear as the thermal state is simultaneously diagonal in the coherent-state and photon-number bases, cf. Eq. \eqref{eq:ThermalState}.
	Conversely, a squeezed state is an example of a state that exhibits quantum coherence in both pictures as demonstrated via our main result.
	It is also worth recalling that a coherent state is a quantum superposition of photon-number states, $|\alpha\rangle=\sum_{n=0}^\infty e^{-|\alpha|^2/2}\alpha^n |n\rangle/\sqrt{n!}$, and a photon-number state can be expanded via coherent states, $|n\rangle=\int d^2\alpha\,\pi^{-1}e^{-|\alpha|^2/2}\alpha^{\ast n} |\alpha\rangle/\sqrt{n!}$, using the identity in the form $\hat 1=\pi^{-1}\int d^2\alpha |\alpha\rangle\langle\alpha|$.
	Both show that the quantum superposition principle is required to describe these states in their complementary bases, representing the fundamental idea of the notion of quantum coherence.

	As a final remark, it is worth emphasizing that $e_\mathrm{wave/part.}\geq0$ means that the outcome of our analysis is \textit{compatible} with the respective picture, wave/particle.
	It does not mean that this description in a classical picture is necessarily true (or the only consistent description) since other measurement scenarios could reveal other forms of inconsistencies, i.e., quantum coherence.
	However, a significant deviation from the derived constraints is sufficient to certify without a doubt that the picture under study is not applicable.
	As it is common in physics, a result cannot be validated, but it can be falsified.

\section{Pseudo-photon-number resolution}\label{sec:dataanalysis}

	To compare our detection with idealized detectors, having a perfect photon-number resolution, we consider a moment-based approach.
	Factorial moments, obtained from the joint click-counting statistics $c_{k_A,k_B}$ measured with time-bin multiplexing detectors (TMDs) which have $D$ detection bins each, take the form \cite{SVA13}
	\begin{equation}
	\begin{split}
		&\mathrm{E}[(k_A)_{m_A}(k_B)_{m_B}]
		=\sum_{k_A,k_B=0}^D (k_A)_{m_A}(k_B)_{m_B} c_{k_A,k_B}
		\\
		=&(D)_{m_A}(D)_{m_B}\langle{:}\left[\hat 1-e^{-\hat n_A/D}\right]^{m_A}\left[\hat 1-e^{-\hat n_B/D}\right]^{m_B}{:}\rangle,
	\end{split}
	\end{equation}
	where the notation $(x)_m=x(x{-}1)\cdots(x{-}m{+}1)=x!/(x{-}m)!$ is used and ${:}\cdots{:}$ denotes the normal ordering \cite{VW06,A12}.
	Note that the above relation applies to ideal TMDs and imperfections, such as a finite quantum efficiency, can be handled by introducing attenuation to the state, rather than the detector.
	In addition, it is worth recalling that from the factorial moments, one can directly reconstruct the moments as well.

	In order to estimate photon-number moments from click-counting detectors, we can apply a Taylor series expansion as long as the photon number distribution is sufficiently bounded (see Ref. \cite{KSVS18} for details),
	\begin{equation}
	\label{eq:ClickApprox}
	\begin{split}
		&\langle{:}\left[\hat 1-e^{-\hat n_A/D}\right]^{m_A}\left[\hat 1-e^{-\hat n_B/D}\right]^{m_B}{:}\rangle
		\\
		\approx&
		\frac{\langle{:}\hat n_A^{m_A}\hat n_B^{m_B}{:}\rangle}{D^{m_A+m_B}}
		-\frac{
			m_A\langle{:}\hat n_A^{m_A+1}\hat n_B^{m_B}{:}\rangle
			+m_B\langle{:}\hat n_A^{m_A}\hat n_B^{m_B+1}{:}\rangle
		}{2D^{m_A+m_B+1}}.
	\end{split}
	\end{equation}
	Note that for applying this approximation, low intensities have to be chosen in our experiment.
	Combining this approximation with the above representation of factorial moments, we find the lowest-order approximation of the factorial moments of the photon number as
	\begin{equation}
		\label{eq:LowestOrderApprox}
		\langle{:}\hat n_A^{m_A}\hat n_B^{m_B}{:}\rangle
		\approx D^{m_A+m_B}
		\overbrace{
			\left[\frac{\mathrm{E}[(k_A)_{m_A}(k_B)_{m_B}]}{(D)_{m_A}(D)_{m_B}}\right]
		}^{\text{\normalsize
			$\stackrel{\text{def.}}{=}M_{m_A,m_B}$
		}}.
	\end{equation}
	Furthermore, the next-order correction [cf. Eq. \eqref{eq:ClickApprox}] provides the systematic error estimate,
	\begin{equation}
	\begin{split}
		&\left|
			\langle{:}\hat n_A^{m_A}\hat n_B^{m_B}{:}\rangle
			-D^{m_A+m_B}M_{m_A,m_B}
		\right|
		\\
		\lessapprox&\frac{
			m_A\langle{:}\hat n_A^{m_A+1}\hat n_B^{m_B}{:}\rangle
			+m_B\langle{:}\hat n_A^{m_A}\hat n_B^{m_B+1}{:}\rangle
		}{2D}
		\\
		\approx&
		\frac{D^{m_A+m_B}}{2}\left[
			m_A M_{m_A+1,m_B}
			+m_B M_{m_A,m_B+1}
		\right],
	\end{split}
	\end{equation}
	where the third line is obtained by applying the approximation in Eq. \eqref{eq:LowestOrderApprox}.

	As final remarks, let us firstly recall that we have \cite{SVA13}
	\begin{equation}
		M_{m_A,m_B}
		=\sum_{k_A,k_B=0}^D \frac{\binom{k_A}{m_A}\binom{k_B}{m_B}}{\binom{D}{m_A}\binom{D}{m_B}} c_{k_A,k_B},
	\end{equation}
	where $\binom{x}{m}=0$ for the binomial coefficients when $x<m$.
	This relation for $M_{m_A,m_B}$ can be directly applied to our data to estimate mean values and random errors.
	Secondly, we can directly estimate photon-number moments from the factorial ones in Eq. \eqref{eq:LowestOrderApprox},
	\begin{equation}
	\begin{split}
		\langle\hat n_A\rangle \approx D M_{1,0},
		\quad
		\langle\hat n_B\rangle \approx D M_{0,1},
		\quad
		\langle\hat n_A\hat n_B\rangle \approx D^2 M_{1,1},
		\\
		\langle\hat n_A^2\rangle \approx D^2 M_{2,0}+ D M_{1,0},
		\text{ and }
		\langle\hat n_B^2\rangle \approx D^2 M_{0,2}+ D M_{0,1},
	\end{split}
	\end{equation}
	using ${:}\hat n{:}=\hat n$ and ${:}\hat n^2{:}=\hat n(\hat n-\hat 1)$.
	Thirdly, we have $D=8$.

	It is also worth mentioning that a standard quadratic error propagation is applied throughout our data analysis.
	For details on the experimental estimation of minimal eigenvalues of a matrix, see, e.g., the Supplemental Material to Ref. \cite{SBVHBAS15}.

\section{Details on the setup}\label{sec:experiement}

\begin{figure}
	\includegraphics[width=\columnwidth]{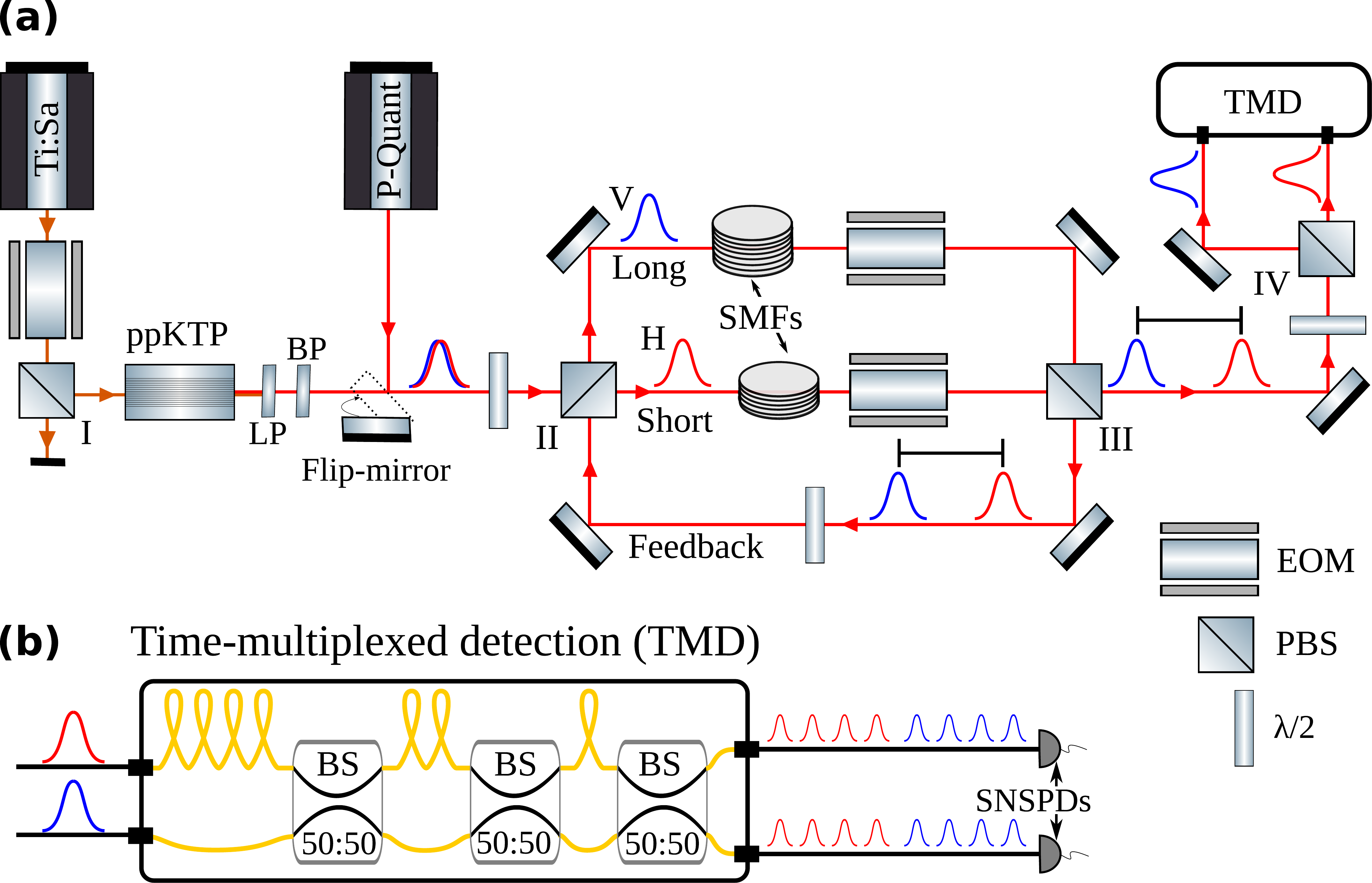}
	\caption{
		Schematic of the setup.
		Ti:Sa: titanium sapphire laser;
		P-Quant: PicoQuant laser;
		ppKTP: periodically poled potassium titanyl phosphate waveguide;
		LP: longpass filter;
		BP: bandpass filter;
		EOM: electro-optic modulator;
		PBS: polarizing beam splitter;
		$\lambda/2$: half-wave plate;
		SMFs: single-mode fibers;
		H,V: horizontal and vertical polarization;
		BS: beam splitter;
		SNSPDs: superconducting nanowire single-photon detectors.
	}\label{fig:setup}
\end{figure}

	A schematic depiction of the experimental setup is shown in Fig. \ref{fig:setup}.
	A titanium sapphire (Ti:Sa) laser delivers pico\-second pulses, centered at $773.5\,\mathrm{nm}$ with a full-width half maximum (FWHM) of $0.3\,\mathrm{nm}$.
	This laser beam, after passing through an elctro-optic modulator (EOM) and a polarizing beam splitter (PBS-I), is coupled into a $2.5\,\mathrm{cm}$ long periodically poled potassium titanyl phosphate (ppKTP) waveguide.
	See \cite{MPEQDBS18} for a detailed characterization of a similar waveguide source.
	The combination of the EOM and the PBS enables down-sampling of the pulse repetition rate (from $76\,\mathrm{MHz}$ to $76\,\mathrm{kHz}$) that is compatible with our time-multiplexing scheme.
	These picosecond pulses pump the \mbox{ppKTP} waveguide that generates two-mode squeezed vacuum (TMSV) via type-II spontaneous parametric down-conversion (SPDC).
	This choice of the pump spectrum and the waveguide length ensure a single Schmidt mode (FWHM $1.2\,\mathrm{nm}$) for the two down-converted polarizations (signal and idler both centered at $1547\,\mathrm{nm}$).
	Moreover, the picosecond SPDC source guarantees a negligible effect of dispersion inside the fiber-loop.
	The pump is filtered out using a longpass filter.
	After passing through a bandpass filter (FWHM $2\,\mathrm{nm}$), the two down-converted modes are sent to a fiber-loop based time-multiplexing setup that acts as a reconfigurable interferometer in the present study; see \cite{NBKSSGPKJS18} for a detailed characterisation.

	In our time-multiplexing part, PBS-II directs horizontal (H) and vertical (V) polarization towards two paths of different fiber lengths while PBS-III redirects them along the same path, thus creating a well-defined delay ($104\,\mathrm{ns}$) between the two polarizations, cf. Fig. \ref{fig:setup}(a).
	By manipulating the polarization of the pulses with two fast-switching EOMs in the two paths, we control whether the pulses are fed back into the loop or are directed to the detection unit.
	This deterministic in- and out-coupling facilitates dynamical reconfiguration of the interferometer. 
	The two polarizations exiting the time-multiplexing loop are symmetrically mixed using the combination of a half-wave plate (HWP) and PBS-IV.

	Finally, the two beam splitter outputs are detected using a two-mode time-multiplexed detection (TMD) \cite{SBVHBAS15} where each pulse is divided into eight subsequent pulses (time bins) that are detected in a time-resolved way with two superconducting nanowire single-photon detectors (SNSPDs), shown in Fig. \ref{fig:setup}(b).
	The separation between the time bins is around $100\,\mathrm{ns}$ that allows to resolve them almost perfectly with our SNSPDs, having a $50\,\mathrm{ns}$ dead time. From the eight-bin click distribution of each pulse, we construct covariance matrices of the two pulses injected into the TMD, cf. Sec. \ref{sec:dataanalysis}.

	In case of measurements with the squeezed vacuum states, a HWP at $22.5^{\circ}$ followed by PBS-II symmetrically mixes the two down-converted polarization modes.
	This creates two separated squeezed states in H and V polarization, propagating along the short and long path of the fiber-loop, respectively.
	The EOMs then switch the polarizations (H to V and vice versa) and sent them into the feedback path, where a HWP at $45^{\circ}$ enforces another polarization flip.
	Thus, the H pulse that traversed through the short path is now redirected towards the long path, while the V light coming from the long path goes to the short path.
	Next, the EOMs perform another polarization switching to ensure that the two polarizations exit the loop.
	This particular design of the time-multiplexing loop ensures that two squeezed states at the output are temporally overlapped.
	Henceforth, the subsequent HWP at $22.5^{\circ}$ and PBS-IV implement symmetric (i.e., 50:50) mixing of the two squeezed states, and the outputs are detected with TMD. 
 
	It is worth noting that the above scheme successively implements two beam splitter operations on a TMSV state---one aimed at producing two separable squeezed states and another one to implement the beam splitter in our interference device.
	In principle, this returns the same TMSV state under ideal conditions.
	Therefore, we additionally analyze the TMSV state from the SPDC source where the two down-converted modes are directly sent to the TMD without any mixing.
	This is achieved simply by removing all the HWPs from the setup and switching-off the EOMs.
	As expected, the results (not presented here) show similar trend as in Fig. 3 of the main text, however, with higher quantum efficiency, $\eta\approx 12\%$, because the fibre-loop network contributes less to the total loss. 

	For the measurement with coherent states, we put the flip-mirror up that couples picosecond light pulses from a laser source (PicoQuant, centered at $1550\,\mathrm{nm}$, FWHM $\sim0.2\,\mathrm{nm}$) into the time-multiplexing loop while blocking the SPDC light.
	We utilize the same loop design as in the squeezed-state-based experiment that now leads to two coherent states of orthogonal polarizations at the output of the loop.
	The two coherent states are symmetrically mixed at the PBS-IV and also detected with TMD.


\begin{thebibliography}{99}
	\bibitem{B28}
		N. Bohr,
		\textit{The Quantum Postulate and the Recent Development of Atomic Theory},
		\href{https://doi.org/10.1038/121580a0}{Nature (London) \textbf{121}, 580 (1928)}.
	\bibitem{AI38}
		A. Einstein and L. Infeld,
		\textit{The Evolution of Physics: The Growth of Ideas from Early Concepts to Relativity and Quanta},
		(\href{https://doi.org/10.1017/S0031819100011670}{Cambridge University Press, Cambridge, UK, 1938}).
	\bibitem{DG28}
		C. Davisson and L. H. Germer,
		\textit{Diffraction of Electrons by a Crystal of Nickel},
		\href{https://doi.org/10.1103/PhysRev.30.705}{Phys. Rev. \textbf{30}, 705 (1927)}.
	\bibitem{KDM77}
		H. J. Kimble, M. Dagenais, and L. Mandel,
		\textit{Photon Antibunching in Resonance Fluorescence},
		\href{https://doi.org/10.1103/PhysRevLett.39.691}{Phys. Rev. Lett. \textbf{39}, 691 (1977)}.
	\bibitem{GRA86}
		P. Grangier, G. Roger, and A. Aspect,
		\textit{Experimental Evidence for a Photon Anticorrelation Effect on a Beam Splitter: A New Light on Single-Photon Interferences},
		\href{https://doi.org/10.1209/0295-5075/1/4/004}{Europhys. Lett. \textbf{1}, 173 (1986)}.
	\bibitem{P00}
		M. Planck,
		\textit{Zur Theorie des Gesetzes der Energieverteilung im Normalspectrum},
		\href{https://doi.org/10.1002/phbl.19480040404}{Verhandlungen der Deutschen Physikalischen Gesellschaft \textbf{2}, 237 (1900)}.
	\bibitem{E05}
		A. Einstein,
		\textit{\"Uber einen die Erzeugung und Verwandlung des Lichts betreffenden heuristischen Gesichtspunkt},
		\href{https://doi.org/10.1002/andp.19053220607}{Annalen der Physik \textbf{17}, 132 (1905)}.
	\bibitem{MW95}
		L. Mandel and E. Wolf,
		\textit{Optical Coherence and Quantum Optics}
		(\href{https://doi.org/10.1017/CBO9781139644105}{Cambridge University Press, Cambridge, UK, 1995}).
	\bibitem{VW06}
		W. Vogel and D.-G. Welsch,
		\textit{Quantum Optics}
		(\href{https://doi.org/10.1002/3527608524}{Wiley-VCH, Weinheim, 2006}).
	\bibitem{A12}
		G. S. Agarwal,
		\textit{Quantum Optics}
		(\href{https://doi.org/10.1017/CBO9781139035170}{Cambridge University Press, Cambridge, 2012})
	\bibitem{KLM01}
		E. Knill, R. Laflamme, and G. J. Milburn,
		\textit{A scheme for efficient quantum computation with linear optics},
		\href{https://doi.org/10.1038/35051009}{Nature (London) \textbf{409}, 46 (2001)}.
	\bibitem{KMNRDM07}
		P. Kok, W. J. Munro, K. Nemoto, T. C. Ralph, J. P. Dowling, and G. J. Milburn,
		\textit{Linear optical quantum computing with photonic qubits},
		\href{https://doi.org/10.1103/RevModPhys.79.135}{Rev. Mod. Phys. \textbf{79}, 135 (2007)}.
	\bibitem{GT07}
		N. Gisin and R. Thew,
		\textit{Quantum communication},
		\href{https://doi.org/10.1038/nphoton.2007.22}{Nat. Photonics \textbf{1}, 165 (2007)}.
	\bibitem{S09}
		J. H. Shapiro,
		\textit{The quantum theory of optical communications},
		\href{https://doi.org/10.1109/JSTQE.2009.2024959}{IEEE J. Sel. Top. Quantum Electron. \textbf{15}, 1547 (2009)}.
	\bibitem{HOM87}
		C. K. Hong, Z. Y. Ou, and L. Mandel,
		\textit{Measurement of subpicosecond time intervals between two photons by interference},
		\href{https://doi.org/10.1103/PhysRevLett.59.2044}{Phys. Rev. Lett. \textbf{59}, 2044 (1987)}.
	\bibitem{B69}
		D. R. Brillinger,
		\textit{The calculation of cumulants via conditioning},
		\href{https://doi.org/10.1007/BF02532246}{Ann. Inst. Stat. Math. \textbf{21}, 215 (1969)}.
	\bibitem{S95}
		M. J. Schervish,
		\textit{Theory of Statistics}
		(\href{https://doi.org/10.1007/978-1-4612-4250-5}{Springer, New York, NY, 1995}).
	\bibitem{B64}
		J. S. Bell,
		\textit{On the Einstein Podolsky Rosen paradox},
		\href{https://doi.org/10.1103/PhysicsPhysiqueFizika.1.195}{Physics \textbf{1},195 (1964)}.
	\bibitem{Supplement}
		See the Supplemental Material, including the Refs. \cite{VW06,A12,SBVHBAS15,SVA13,KSVS18,MPEQDBS18}, for technical details on the quantum-mechanical treatment, including losses, as well as our data processing and additional experimental descriptions.
	\bibitem{NBKSSGPKJS18}
		T. Nitsche, S. Barkhofen, R. Kruse, L. Sansoni, M. \v{S}tefa\v{n}\'{a}k, A. G\'{a}bris, V. Poto\v{c}ek, T. Kiss, I. Jex, and C. Silberhorn,
		\textit{Probing measurement-induced effects in quantum walks via recurrence},
		\href{https://doi.org/10.1126/sciadv.aar6444}{Sci. Adv. \textbf{4}, eaar6444 (2018)}.
	\bibitem{SBVHBAS15}
		J. Sperling, M. Bohmann, W. Vogel, G. Harder, B. Brecht, V. Ansari, and C. Silberhorn,
		\textit{Uncovering Quantum Correlations with Time-Multiplexed Click Detection},
		\href{https://doi.org/10.1103/PhysRevLett.115.023601}{Phys. Rev. Lett. \textbf{115}, 023601 (2015)}.
	\bibitem{BBCH16}
		E. Bagan, J. A. Bergou, S. S. Cottrell, and M. Hillery,
		\textit{Relations between Coherence and Path Information},
		\href{https://doi.org/10.1103/PhysRevLett.116.160406}{Phys. Rev. Lett. \textbf{116}, 160406 (2016)}.
	\bibitem{RPMAXSFS17}
		A. S. Rab, E. Polino, Z.-X. Man, N. B. An, Y.-J. Xia, N. Spagnolo, R. Lo Franco, and F. Sciarrino,
		\textit{Entanglement of photons in their dual wave-particle nature},
		\href{https://doi.org/10.1038/s41467-017-01058-6}{Nat. Commun. \textbf{8}, 915 (2017)}.  
	\bibitem{YHZZXLG18}
		Y. Yuan, Z. Hou, Y.-Y. Zhao, H.-S. Zhong, G.-Y. Xiang, C.-F. Li, and G.-C. Guo,
		\textit{Experimental demonstration of wave-particle duality relation based on coherence measure},
		\href{https://doi.org/10.1364/OE.26.004470}{Opt. Express \textbf{26}, 4470 (2018)}.
	\bibitem{DCPDS18}
		S. Das, I. Chakrabarty, A. K. Pati, Aditi S. De, and U. Sen,
		\textit{Quantifying the particle aspect of quantum systems},
		\href{https://arxiv.org/abs/1812.08656}{arXiv:1812.08656}.
	\bibitem{QMLYN19}
		W. Qin, A. Miranowicz, G. Long, J. Q. You, and F. Nori,
		\textit{Proposal to test quantum wave-particle superposition on massive mechanical resonators},
		\href{https://doi.org/10.1038/s41534-019-0172-9}{npj Quantum Inf. \textbf{5}, 58 (2019)}.
	\bibitem{CBB19}
		E. G. Carnio, H.-P. Breuer, and A. Buchleitner,
		\textit{Wave-Particle Duality in Complex Quantum Systems},
		\href{https://doi.org/10.1021/acs.jpclett.9b00676}{J. Phys. Chem. Lett. \textbf{10}, 2121 (2019)}.
	\bibitem{QA19}
		X.-F. Qian, G. S. Agarwal
		\textit{Are Quantum Objects Born with Duality?}
		\href{https://arxiv.org/abs/1901.07595}{arXiv:1901.07595}.
	\bibitem{QKMSVW19}
		X.-F. Qian, K. Konthasinghe, K. Manikandan, D. Spiecker, A. N. Vamivakas, and J. H. Eberly,
		\textit{Can Quantum Duality be Turned Off?},
		\href{https://arxiv.org/abs/1907.01718}{arXiv:1907.01718}.
	\bibitem{GG02}
		F. Grosshans and P. Grangier,
		\textit{Continuous Variable Quantum Cryptography Using Coherent States},
		\href{https://doi.org/10.1103/PhysRevLett.88.057902}{Phys. Rev. Lett. \textbf{88}, 057902 (2002)}.
	\bibitem{GAWBCG03}
		F. Grosshans, G. Van Assche, J. Wenger, R. Brouri, N. J. Cerf, and P. Grangier 
		\textit{Quantum key distribution using gaussian-modulated coherent states},
		\href{https://doi.org/10.1038/nature01289}{Nature (London) \textbf{421}, 238 (2003)}.
	\bibitem{G63}
		R. J. Glauber,
		\textit{Coherent and incoherent states of the radiation field},
		\href{https://doi.org/10.1103/PhysRev.131.2766}{Phys. Rev. \textbf{131}, 2766 (1963)}.
	\bibitem{S63}
		E. C. G. Sudarshan,
		\textit{Equivalence of semiclassical and quantum mechanical descriptions of statistical light beams},
		\href{https://doi.org/10.1103/PhysRevLett.10.277}{Phys. Rev. Lett. \textbf{10}, 277 (1963)}.
	\bibitem{TG65}
		U. M. Titulaer and R. J. Glauber,
		\textit{Correlation functions for coherent fields},
		\href{https://doi.org/10.1103/PhysRev.140.B676}{Phys. Rev. \textbf{140}, B676 (1965)}.
	\bibitem{L86}
		L. Mandel,
		\textit{Non-Classical States of the Electromagnetic Field},
		\href{https://doi.org/10.1088/0031-8949/1986/T12/005}{Phys. Scr. \textbf{T12}, 34 (1986)}.
	\bibitem{SAP17}
		A. Streltsov, G. Adesso, and M. B. Plenio,
		\textit{Quantum coherence as a resource},
		\href{https://doi.org/10.1103/RevModPhys.89.041003}{Rev. Mod. Phys. \textbf{89}, 041003 (2017)}.
	\bibitem{CG18}
		E. Chitambar and G. Gour,
		\textit{Quantum Resource Theories},
		\href{https://doi.org/10.1103/RevModPhys.91.025001}{Rev. Mod. Phys. \textbf{91}, 025001 (2019)}.
	\bibitem{VS14}
		W. Vogel and J. Sperling,
		\textit{Unified quantification of nonclassicality and entanglement},
		\href{https://doi.org/10.1103/PhysRevA.89.052302}{Phys. Rev. A \textbf{89}, 052302 (2014)}.
	\bibitem{KSP16}
		N. Killoran, F. E. S. Steinhoff, and M. B. Plenio,
		\textit{Converting Nonclassicality into Entanglement},
		\href{https://doi.org/10.1103/PhysRevLett.116.080402}{Phys. Rev. Lett. \textbf{116}, 080402 (2016)}.
	\bibitem{FP12}
		A. Ferraro and M. G. A. Paris,
		\textit{Nonclassicality Criteria from Phase-Space Representations and Information-Theoretical Constraints Are Maximally Inequivalent},
		\href{https://doi.org/10.1103/PhysRevLett.108.260403}{Phys. Rev. Lett. \textbf{108}, 260403 (2012)}.
	\bibitem{ASV13}
		E. Agudelo, J. Sperling, and W. Vogel,
		\textit{Quasiprobabilities for multipartite quantum correlations of light},
		\href{https://doi.org/10.1103/PhysRevA.87.033811}{Phys. Rev. A \textbf{87}, 033811 (2013)}.
	\bibitem{Setal17}
		J. Sperling, W. R. Clements, A. Eckstein, M. Moore, J. J. Renema, W. S. Kolthammer, S. W. Nam, A. Lita, T. Gerrits, W. Vogel, G. S. Agarwal, and I. A. Walmsley,
		\textit{Detector-Independent Verification of Quantum Light},
		\href{https://doi.org/10.1103/PhysRevLett.118.163602}{Phys. Rev. Lett. \textbf{118}, 163602 (2017)}.
	\bibitem{SVA13}
		J. Sperling, W. Vogel, and G. S. Agarwal,
		\textit{Correlation measurements with on-off detectors},
		\href{https://doi.org/10.1103/PhysRevA.88.043821}{Phys. Rev. A \textbf{88}, 043821 (2013)}.
	\bibitem{KSVS18}
		O. P. Kovalenko, J. Sperling, W. Vogel, and A. A. Semenov,
		\textit{Geometrical picture of photocounting measurements},
		\href{https://doi.org/10.1103/PhysRevA.97.023845}{Phys. Rev. A \textbf{97}, 023845 (2018)}.
	\bibitem{MPEQDBS18}
		E. Meyer-Scott, N. Prasannan, C. Eigner, V. Quiring, J. M. Donohue, S. Barkhofen, and C. Silberhorn,
		\textit{High-performance source of spectrally pure, polarization entangled photon pairs based on hybrid integrated-bulk optics},
		\href{https://doi.org/10.1364/OE.26.032475}{Opt. Express \textbf{26}, 32475 (2018)}.
\end{thebibliography}
\end{document}